# CALCULATION OF THE ELASTIC p$^6$He AND p$^8$He SCATTERING IN GLAUBER APPROXIMATION

ELENA IBRAEVA,[1,*] ONLASYN IMAMBEKOV[1,2] and ALBERT DZHAZAIROV-KAHRAMANOV[1,3,†]

[1] *Institute of nuclear physics RK, 050032, str. Ibragimova 1, Almaty, Kazakhstan*
[2] *Al-Farabi Kazakh National University, 050040, av. Al-Farabi 71, Almaty, Kazakhstan*
[3] *V. G. Fessenkov Astrophysical Institute "NCSRT" NSA RK, 050020, Observatory 23, Kamenskoe plato, Almaty, Kazakhstan*
[*] *ibraeva.elena@gmail.com*
[†] *albert-j@yandex.ru*

**Abstract:** The series of calculations of differential cross sections and analyzing powers ($A_y$) of the elastic proton scattering on isotopes $^6$He and $^8$He was presented in the framework of the theory of Glauber multiple diffraction scattering. The three-body $\alpha-n-n$ and harmonic oscillator wave functions were used for $^6$He, for $^8$He – the function of density distribution in LSSM. For $p^6$He scattering the expansion of the Glauber operator into a series of multiple scattering is written in the form consistent with the picture of weakly bound clusters in halo-nuclei, for $p^8$He scattering the members of single- and double collisions are included in the operator. We have done the comparison of our results with available experimental data (including new data of $A_y$ for $p^6$He scattering at the $E$ = 71 MeV/nucleon) and with the calculations results in the other approaches (high-energy, optical model, RMF). Also, the comparison of the characteristics of two isotopes $^6$He and $^8$He was made, which showed that they correlate quite well with each other.

## 1. Introduction

Capabilities for studies of nuclear matter have greatly been expanded with beams of radioactive nuclei. Measurements of differential and total cross sections for proton scattering from these nuclei in the inverse kinematics provide important information about their structure: irregularities in neutron and proton densities (halo), new deformation regions and new type of collective excitations at low energies (soft dipole resonance), non-regularities in shell population, etc.

The experiments on elastic protons scattering on He isotopes in inverse kinematics at high energy (700 MeV/nucleon) were held in GSI by G.D. Alkhazov[1,2] using the hydrogen-filled ionization chamber IKAR served simultaneously as a gas target and a detector for the recoil protons. These experiments at small momentum transfers ($t$ < 0.05 (GeV/c)$^2$) provided valuable information on nuclear size and $^{6,8}$He mass density distribution. However at higher momentum transfer (especially in the diffraction minima), the differential cross section (DCS) should be more sensitive to the details of nuclei structure. Therefore the experimental studies have been focused on cross-sections measurements at higher momentum transfer, which was done in Refs. 3, 4, where $t$ values are measured up to 0.125 (GeV/c)$^2$ Ref. 4 and 0.20 (GeV/c)$^2$ Ref. 3.

The main difference from the previous experiment[1,2] – is the use of liquid hydrogen target in combination with protons recoil detector instead of gas target.

The experiments at intermediate energies (up to 100 MeV/nucleon) are made on RARF (RIKEN Acclerator Research Facility). The first measurements of DCS of the $p^8$He scattering are published in Ref. 5 at $E = 72$ MeV/nucleon in the angular range $\theta \sim 26-64°$; $p^6$He scattering at $E = 71$ MeV/nucleon in the angular range $\theta \sim 20-48°$ are published in Ref. 6. Further measurements have expanded this area to $\theta \sim 87°$.[7–9]

Creation of a solid polarized proton target has become a new achievement in the experiments with radioactive beams.[7–9] Gas target, due to its low density combined with low intensity of radioactive beams, makes the experiment very difficult. The main difficulty was to create a solid polarized proton target, which could work in a weak magnetic field. The way to create such target was found, based on a new principle, which does not depend on the strength of magnetic field. The protons in the target are polarized by the polarized electrons transferable in photo-excited triplet states from pentacene molecules. The value of electron polarization is 73% and is independent neither of the magnetic field strength nor the material temperature. This target can be operated in a weak magnetic field ($\sim 0.1\ T$), and at a sufficiently high temperature ($\sim 100\ K$). The vector analyzing power in the elastic $p^6$He scattering was measured at 71 MeV/nucleon[7,8] with the help of solid polarized proton target.

The theoretical study of elastic scattering at intermediate energies is usually held within the optical model or the Glauber diffraction theory.[10] The latter makes it possible to describe the proton-nucleus scattering without free parameters (wave function (WF) is calculated independently with the fixed potentials of intercluster interactions, the parameters of elementary $pN$-amplitudes are found from the experiments on nucleon-nucleon scattering) and to extract information directly from the measured values. Taking into account the spin component in the $pN$-interaction makes it possible to calculate the scattering properties such as polarization, analyzing power, spin-rotation function and etc., depending on relative orientation of the spins of the colliding particles. It is well known that the polarization observables constitute the sensitive probe of nuclear structure and interaction mechanism. The spin-orbit interaction in the scattering of the polarized proton beams on stable targets is measured in numerous experiments.

Currently, the measurements of spin asymmetry are used to study unstable nuclei. The nuclei, located near neutron stability drip-line, have a substantial structure, such as halo or skin (excess of neutrons at nucleus surface). Since the spin-orbit coupling in the nucleus is essentially a surface effect and excess neutron distribution is observed at the surface of the nucleus, it is interesting to see how the halo (or skin) affects the spin asymmetry (i.e., vector analyzing power $A_y$) in the elastic scattering of protons.

Although, until recently, there was no experiments to measure the polarization characteristics for radioactive nuclei, but there are theoretical predictions of these characteristics (in particular, analyzing power) on isotopes He, Li, Be, C and others.[11-15] Thus, the Ref. 11 provides DCS calculation and $A_y$ for $p^9$Li-elastic scattering at $E = 60$ MeV/nucleon in the optical model with two effective interactions JLM (Jeukenne-Lejeune-Maxauh) and with the Paris-Hamburg nonlocal potential. It is shown that polarization characteristics can be radically different from each other even with relatively small variations of calculation. Thus, the analyzing powers, calculated with different potentials of interaction, are opposite at certain angles.

The Ref. 12 calculates (and compares with experimental data) DCSs and $A_y$ of elastic scattering of 65 MeV protons on nuclei from $^6$Li to $^{238}$U. The calculation is performed in the optical model with nonlocal potentials obtained from the folding of complex effective interaction potential with single-particle density matrix for each target. The Large Scale Shell Model (LSSM) was used for light nuclei. The effective interaction is obtained from the solution of the Lippmann-Schwinger and Bruckner-Bethe-Goldstoun equations with Paris potential. The calculations were compared with two densities: with and without the effects of nuclear medium. The differences in DCSs description are insufficient, while $A_y$ differs significantly throughout the angular range from 0 to 80°.

The Ref. 13 calculates the DCSs and $A_y$ for the p$^8$He elastic scattering at $E$ = 72 and 200 MeV/nucleon in the single scattering impulse approximation in the expansion of multiple scattering of the optical potential (OP) in KMT (Kerman-McManus-Thaler) formulation. In the later paper Ref. 14 DCS and $A_y$ for $p^{4,6,8}$He scattering at 297 MeV are calculated in the same approximation. The structure of $^6$He is represented by three-body $\alpha-n-n$ and oscillatory WFs. The behavior of $A_y$ with both WFs matches at small angles, although the distribution of mass density of valence neutrons is different. This fact indicates that the spin-orbital contribution of halo valence neutrons is very small due to short distribution scope of their material density, when it is turned off with the spin-orbit component of the nucleon-nucleon scattering amplitude (which is close to zero at low momentum transfer), which gives a negligible contribution to the total spin-orbit force.

The Ref. 15 represents the prediction of $A_y$ in the full $t$-folding OP model for $p^{6,8}$He scattering at $E$ = 66–100 MeV. In this model, the amplitude of the proton-nucleus scattering is constructed by folding amplitudes of free nucleon-nucleon scattering ($t$-matrix) with the off-shell density matrixes.

However, none of the calculations performed earlier, do not describe the experimental data obtained in Refs. 8, 9. For example, in Ref. 15 the theory predicts large positive values of the analyzing power (up to 0.8 at $\theta$ = 56°), which is clearly not consistent with the experimental data where the maximum positive value of 0.24 is achieved at $\theta$ = 37°, the minimum is -0.27 at $\theta$ = 74°.[8] The calculations of the analyzing power, presented in Ref. 8 in the optical model with WFs, calculated in the Woods-Saxon potential and in the potential of harmonic oscillator, have shown that the approximate agreement with the experiment (up to $\theta$ < 52°) is only for $^6$He WF in the Woods-Saxon potential with halo structure. The calculations, made in g-matrix folding model, reproduce DCS data in the entire angular range (90° > $\theta$ > 20°) only when $^6$He has a distinct $\alpha$-core, whereas $A_y$ agrees with the experiment only up to $\theta$ < 55°.

Continuation of Ref. 8 was the paper Ref. 9 that represents the parts of the experiment in more details and theoretical calculations. The joint analysis of $A_y$ and DCS is conducted in three models: the cluster folding (with the $\alpha-n-n$ cluster structure of $^6$He), the nucleon folding (with the non-clustered $2p + 4n$-structure of $^6$He), and the fully microscopic model with nonlocal OPs, with three sets of single-particle WFs. This analysis showed that the DCS described well enough in all models (especially when $\theta$ < 70°). The DCS sensitivity to the details of $^6$He WF structure is found near the minimum of the cross section ($\theta \sim 53°$) and for large angles ($\theta$ > 70°). The fitting of the cross section is made by variation of OP parameters (mainly by increasing core and halo diffusion). It is more difficult to describe the analyzing power. The experimental data are not reproduced correctly in any model, especially the region 70° > $\theta$ > 40° close to zero of $A_y$ values, as well as the last point at $\theta \sim 75°$. However, the description of the polarization characteristics includes the successful calculation that at least qualitatively reproduces the data. The Ref. 9 contains such calculation with the phenomenological OP (set B) and $\alpha-n-n$ cluster-folding potential.

At the same time, on the basis of the performed calculations it was possible to make the important conclusions about the OP form and the dominant interactions. Thus, the analysis in the optical model showed that the spin-orbit potential for $^6$He is characterized by a shallow and long-ranged shape, reflecting the diffuse density of $^6$He.[9] The calculation with nucleon folding-potential most inadequately reproduces $A_y$, whereas the $\alpha-n-n$-cluster folding gives reasonable agreement with the data. This demonstrates the importance of taking into account of $\alpha$-clustering in the $p^6$He elastic scattering description.

Also the $A_y$ for $p^6$He-scattering is described unsatisfactorily in the recent paper,[16] where the calculations are made in the framework of optical model (with single-particle and cluster OP). This paper derives a non-local OP in first-order in the Watson multiple-scattering expansion, which allows us to separate treatment of proton and neutron contributions to the structure and also naturally take into account the contributions of $\alpha$-core and two neutrons. The matrix of $^6$He density is calculated according to COSMA (Cluster-Orbital-Shell-Model Approximation) program. The calculation with single-particle OP, as well as the calculation based on the cluster model leads to the $A_y$ positive for all angles, which is inconsistent with the experiment. Only the calculation with the cluster OP, where the potential for $\alpha$-core is calculated for the case of $\alpha$-particle with the modified NN $t$-matrix, leads to the negative $A_y$ at large angles ($\theta > 60°$) and correctly predicts the position of the last two angular measurements.

All available experimental data on the elastic scattering of protons from isotopes $^{4,6,8}$He at $E = 71$ MeV/nucleon were analyzed in the latest work.[17] The analyze of DSCs and $A_y$ was done in the eikonal approximation, Glauber approximation of the single scattering and using the folding potential in optical model with taking into account spin-orbit interaction. The WFs of isotopes $^{4,6,8}$He were calculated by the variational Monte-Carlo method for nonrelativistic Hamiltonian consisting of the AV$_{1s}$ two-nucleon potential and the U-IX three-nucleon potential. Results of analysis show that for all nuclei the best possible fit with experiment reaches at the first-order of Glauber approximation. The Pauli principle effect, which slightly overpowers the p-He interaction that decreases DSC improving agreement with the experiment, was studied. It was noted that none of the carried out calculations do not describe $A_y$ in the whole measured angular range ($\theta \sim 35-75°$).

The fact that all model calculations correctly reproduce the data for DCS while $A_y$ is very different even for minor variations in the calculation ("$A_y$-problem", Ref. 16), according to the authors of Ref. 9: "This may indicate limitation of the structure model and/or contribution of unaccounted reaction mechanisms that influence the larger momentum transfer results".

This paper presents a series of DCSs and $A_y$ calculations for proton elastic scattering from $^6$He and $^8$He isotopes, in the Glauber multiple diffraction scattering theory.[10] The three-body $\alpha-n-n$ WF[18,19] was used for $^6$He calculated with realistic potentials of intercluster interactions and shell one[20] with WF in the potential of harmonic oscillator. For $^8$He – the function of density distribution in LSSM.[21] For $p^6$He scattering the expansion of the Glauber operator into a series of multiple-scattering is written in the form consistent with the picture of weakly bound clusters in halo-nuclei, for $p^8$He scattering the members of single- and double collisions are included in the operator. The construction of $^6$He WF in the $\alpha-n-n$-model, the choice of inter-cluster potentials of interactions and the method of calculating matrix elements in the Glauber theory are described in details in our previous studies.[22,23] In this paper we compare our calculations of $A_y$ with new experimental data for $p^6$He scattering at $E = 71$ MeV/nucleon Refs. 7−9 and different approaches to estimation of DCS and $A_y$ to determine the validity of using different models. Also, the comparison of the characteristics of two isotopes $^6$He and $^8$He was made, which showed that they correlate quite well with each other.

## 2. Geometrical configuration of $^6$He and $^8$He nuclei

The $^6$He and $^8$He nuclei are neutron-excess β-radioactive with the dominant structure $α + 2n$ and $α + 4n$. The facts justifying the $α−n−n$-model of $^6$He and the $α2n2n$-model of $^8$He are the low binding energy in the channel $α2n$ ($E_{α-2n} = 0.973$ MeV, Ref. 24) and $α4n$ ($E_{α-4n} = 3.1$ MeV, Ref. 24), spectroscopic factor of $α2n$ and $α4n$ channels close to one, and so the measurement of the DCSs of high-energy proton scattering on $^6$He and $^8$He in inverse kinematics,[1-9] which gave the evidence of the well-defined clustering on α-partial core and valence nucleons, which form halo or skin, and the radius of which equals about 0.9 fm.[25,26] The analysis of DCSs data for $p^6$He and $p^8$He-scattering carried out in full microscopic folding-model with WF in (0+2+4) $\hbar\omega$ model space (LSSM) concluded that $^6$He is a typical halo nucleus, while $^8$He is defined as non-halo nucleus, but having the skin.[21] In Ref. 27 has shown that the neutron halo of $^6$He manifests itself via a more than 30% enhancement of the point-neutron radius with respect to the point-proton radius.

The presence of two configurations: dineutron and cigar-like in $^6$He nucleus is established in Ref. 28, 29 in calculation of the $^6$He+$^4$He elastic scattering and in the reaction of two-neutrons transfer $^6$He + $^4$He → $^4$He + $^6$He at $E_{lab.} = 151^{28}$ and $19.6^{29}$ MeV. Meanwhile, each of them has halo structure radii of $4.4^{29}$ fm for dineutron and $4.18^{29}$ fm for cigar-like configuration.

The calculation of three-cluster correlation function in the RGM algebraic version Ref. 30 for $^6$He and $^8$He nuclei confirmed the presence of two WF configurations, called by the authors as triangular (similar to dineutron) and linear (similar to cigar-like.) The calculated configurations of relative position of clusters for $^6$He are compared with the configuration for $^8$He in $α2n2n$-model. It was found that the dominant configuration in $^8$He forms an equilateral triangle with an angle close to the right, the top of which is α-particle, and in the base − dineutron clusters, i.e., there is some intermediate configuration between dineutron and cigar configuration, similar to the configuration of the $P$-state of $^6$He. The difference in geometry is explained by the Pauli principle: $^8$He contains the effective repulsion between dineutron clusters, making them lie on the opposite sides from the α-particle core, in $^6$He the neutrons with opposite spins in the presence of massive α-core are combined in a compact dineutron (its rms radius in the nucleus is 2.52 fm, which is less than the radius of free deuteron by 0.17 fm).

The analysis of geometric form of $^6$He states in $α−n−n$-model[18,19] shows that the $S$-state includes two geometric configurations: dineutron $α+(2n)$ and cigar-like $(n+α+n)$ with α-particle between two neutrons. These configurations are characterized by the following intra-nuclear distances: $α+(2n) – r = 1.7$ fm, $R = 3$ fm (for comparison in Ref. 28 $r = 2$ fm, $R = 3$ fm), $(n+α+n) – r = 4$ fm, $R = 1$ fm (for comparison in Ref. 28 $r = 4.5$ fm, $R = 1.8$ fm), where $r$, $R$ − are the average distances between two neutrons and between the center mass of two neutrons and α-particle. Due to the fact that $r = 1.7$ fm in $α+(2n)$ configuration it follows that dineutron cluster in the core is strongly compressed in comparison with free deuteron, the radius of which equals 2.69 fm. The configuration of the P-state is close to the equilateral triangle with $r = 2.3$ fm, $R = 1.8$ fm, which revolves around the common center mass of nucleus.

It was obtained in the recent paper,[31] in the framework of full antisymmetrized microscopic model for light two-neutrons halo nuclei, that the dineutron peak for $^6$He is located at $r = 1.93$ fm and $R = 2.63$ fm, cigar-like is at $r = 3.82$ fm and $R = 1.03$ fm. In the $^6$He three-body model it was found that the peaks slightly shifted to larger radii: dineutron peak is at $r = 2.08$ fm and $R = 2.88$ fm, cigar-like is at $r = 4.18$ fm and $R = 1.08$ fm. In both models the probability of dineutron and cigar-like configurations are estimated at 60% and 40%, respectively.

## 3. Brief formalilsm

### 3.1. *The calculation of the central part of the $p^6$He scattering amplitude*

According to the Glauber theory of multiple-scattering[10] the proton elastic scattering amplitude on the compound nucleus with mass $A$ can be written as the integral over the impact parameter $\boldsymbol{\rho}$:

$$M_{if}(\mathbf{q}) = \sum_{M_J M_J'} \frac{ik}{2\pi} \int d\boldsymbol{\rho} \exp(i\mathbf{q}\boldsymbol{\rho}) \langle \Psi_i^{JM_J} | \Omega | \Psi_f^{JM_J'} \rangle, \quad (1)$$

where $\mathbf{q}, \boldsymbol{\rho}$ are the two-dimensional vectors lying in the plane perpendicular to the direction of the incident beam; $\langle \Psi_i^{JM_J} | \Omega | \Psi_f^{JM_J'} \rangle$ is the matrix element of the transition from the initial $\Psi_i^{JM_J}$ to the final $\Psi_f^{JM_J'}$ state of the nucleus under effects of the operator $\Omega$, in case of elastic scattering $\Psi_i^{JM_J} = \Psi_f^{JM_J'}$; $\mathbf{k}, \mathbf{k}'$ are the momenta of projectile and ejectile particles in the center-of-mass system (cms); $q = |\mathbf{k} - \mathbf{k}'| = 2k\sin\frac{\theta}{2}$ is the momentum transfer; $\theta$ is the scattering angle.

Operator $\Omega$ is written as a series of a multiple-scattering:

$$\Omega = 1 - \prod_{\nu=1}^{A}(1-\omega_\nu(\boldsymbol{\rho}-\boldsymbol{\rho}_\nu)) = \sum_{\nu=1}^{A}\omega_\nu - \sum_{\nu<\mu}\omega_\nu\omega_\mu + \sum_{\nu<\mu<\eta}\omega_\nu\omega_\mu\omega_\eta + ...(-1)^{A-1}\omega_1\omega_2...\omega_A, \quad (2)$$

where the first term is responsible for single collisions, the second is for double collisions and etc. to the last term, responsible for $A$-multiple collisions, $\boldsymbol{\rho}_\nu$ is the two-dimensional analog of three-dimensional single-particle nucleons coordinates ($\mathbf{r}_\nu$).

For certainty we suggest to calculate the matrix element for $^6$He nucleus presented in $\alpha-n-n$-model. Let's rewrite the operator of Eq. (2) in the alternative form, based on the fact that the scattering occurs on $\alpha$-particle and two neutrons making up the $^6$He nucleus:

$$\Omega = \Omega_\alpha + \Omega_n + \Omega_n - \Omega_\alpha\Omega_n - \Omega_\alpha\Omega_n - \Omega_n\Omega_n + \Omega_\alpha\Omega_n\Omega_n, \quad (3)$$

where each operator is expressed by the profile functions $\omega_\nu$ as follows:

$$\Omega_\alpha = \omega_\nu(\boldsymbol{\rho}-\mathbf{R}_\alpha) = \frac{1}{2\pi i k}\int d\mathbf{q}\exp(-i\mathbf{q}(\boldsymbol{\rho}-\mathbf{R}_\alpha))f_{p\alpha}(q), \tag{4}$$

$$\Omega_n = \omega_n(\boldsymbol{\rho}-\boldsymbol{\rho}_\nu) = \frac{1}{2\pi i k}\int d\mathbf{q}\exp(-i\mathbf{q}(\boldsymbol{\rho}-\boldsymbol{\rho}_\nu))f_{pn}(q). \tag{5}$$

The above formulas of Eqs. (3), (4) show that the α-particle is considered unstructured and scattering is occurred on it as on a single particle. The central part of the elementary amplitude $f^c_{pn}(q)$ is recorded in a standard way:

$$f^c_{pn}(q) = \frac{k\sigma^c_{pn}}{4\pi}(i+\varepsilon^c_{pn})\exp\left(-\frac{\beta^c_{pn}q^2}{2}\right), \tag{6}$$

where $\sigma^c_{pn}$ parameters is the total cross section of proton scattering on the nucleon, $\varepsilon^c_{pn}$ is the ratio of the real part of the amplitude to the imaginary part, $\beta^c_{pn}$ is the slope parameter of the amplitude cone.

The $f^c_{p\alpha}(q)$ is similarly recorded using the replacement of index $n$ to $\alpha$. The parameters of the elementary amplitude $\sigma^c_{pn}$, $\varepsilon^c_{pn}$, $\beta^c_{pn}$ are the input parameters of the theory, but they are determined from the independent experiments. The parameters of pn-amplitude at $E = 0.7$ and 0.07 GeV/nucleon are taken the same as in.[32] The parameters of pα-amplitude are taken from Ref. 33, 34.

We write the WF of $^6$He with the total angular momentum $J$ (for the ground state $J^\pi = 0^+$, $S = 0$) and its projection $M_J$ in the α−n−n-model:[18]

$$\Psi^{JM_J}_{i,f} = \Psi_\alpha(\mathbf{R}_\alpha)\varphi_{n1}(\mathbf{r}_1)\varphi_{n2}(\mathbf{r}_2)\sum_{\lambda l LS}\Psi^{JM_J}_{\lambda l LS}(\mathbf{r},\mathbf{R}), \tag{7}$$

where $\Psi_\alpha(\mathbf{R}_\alpha)$, $\varphi_{n1,2}(\mathbf{r}_{1,2})$, $\Psi^{JM_J}_{\lambda l LS}(\mathbf{r},\mathbf{R})$ are the WFs of α-particle, neutron (n) and relative motion in the Jacobi coordinates.

**Table.** Interaction potentials, considered in configuration of WFs of $^6$He in α−n−n-model.

| Potential | | | | Model 1, Ref. 18 | Model 2, Ref. 19 |
|---|---|---|---|---|---|
| n-n | | | | Reid with soft core (RSC) | RSC |
| α-n | | | | Sack-Biedenharn-Breit (SBB) | Potential split in orbital-angular momentum parity |
| Configuration | | | | Configuration weight (P) | |
| λ | l | L | S | | |
| 0 | 0 | 0 | 0 | 0.957 | 0.869 |
| 1 | 1 | 1 | 1 | 0.043 | 0.298 |

The weights of $^6$He configurations are shown in table. As it is seen from table, these two components give the maximum contribution in $\alpha-n-n$-model, which we have used in our calculation:

$$\Psi^{JM_J}_{\lambda lLS}(\mathbf{r},\mathbf{R}) = \Psi^{JM_J}_{0000}(\mathbf{r},\mathbf{R}) + \Psi^{JM_J}_{1111}(\mathbf{r},\mathbf{R}), \tag{8}$$

where

$$\Psi^{JM_J}_{\lambda l0LS}(\mathbf{r},\mathbf{R}) = \Psi^{JM_J}_{0000}(\mathbf{r},\mathbf{R}) = \frac{1}{4\pi}\sum_{i,j} C^{00}_{ij}\exp(-\alpha_i \mathbf{r}^2 - \beta_j \mathbf{R}^2) \tag{9}$$

$$\Psi^{JM_J}_{\lambda lLS}(\mathbf{r},\mathbf{R}) =$$
$$= \Psi^{JM_J}_{1111}(\mathbf{r},\mathbf{R}) = \sum_{m\mu M_L M_S}\langle 1m1\mu|1M_L\rangle\langle 1M_L 1M_S|JM_J\rangle Y_{1m}(\mathbf{R})Y_{1\mu}(\mathbf{r})\sum_{i,j} C^{11}_{ij}\cdot r\cdot R\cdot\exp(-\alpha_i \mathbf{r}^2 - \beta_j \mathbf{R}^2) \tag{10}$$

To calculate the matrix element of Eq. (1), we must substitute the WF of Eq. (7) and the operator of Eq. (3) for it, and integrate by all variables that affect the integrand. However, firstly it is required to transfer from the single-particle coordinates to relative Jacobi coordinates that affect the $\Psi^{JM_J}(\mathbf{r},\mathbf{R})$ WF and $\Omega$ operator. The relationship between single-particle and relative coordinates for $^6$He is as follows:

$$\mathbf{r}_1 = \frac{2}{3}\mathbf{R} + \frac{1}{2}\mathbf{r} + \mathbf{R}_6, \quad \mathbf{r}_2 = \frac{2}{3}\mathbf{R} - \frac{1}{2}\mathbf{r} + \mathbf{R}_6, \quad \mathbf{r}_3 = \mathbf{R}_6 - \frac{1}{3}\mathbf{R}, \quad \mathbf{R}_6 = \frac{1}{6}\sum_{i=1}^{6}\mathbf{r}_i. \tag{11}$$

After substitution of elementary amplitude of Eq. (6) in Eq. (5), we integrate by $d\mathbf{q}$:

$$\Omega_n = \omega_n(\mathbf{\rho}-\mathbf{\rho}_\nu) = F_n\exp(-(\mathbf{\rho}-\mathbf{\rho}_\nu)\eta_n), \tag{12}$$

where

$$F_n = \frac{\sigma^c_{pn}}{4\pi(\beta^c_{pn})^2}(1-i\varepsilon^c_{pn}), \quad \eta_n = \frac{1}{2(\beta^c_{pn})^2}. \tag{13}$$

Similarly for $\Omega_\alpha$, with replacement of index $n$ to $\alpha$.

After that, upon moving from the single-particle coordinates to the relative ones and making some transformations, we can write the operator of Eq. (3) in the following form:

$$\Omega = \sum_{k=1}^{7} g_k\exp(-a_k\mathbf{\rho}^2 - b_k\mathbf{R}^2 - c_k\mathbf{r}^2 + d_k\mathbf{\rho}\mathbf{R} + e_k\mathbf{\rho}\mathbf{r} + h_k\mathbf{R}\mathbf{r}), \tag{14}$$

where

$$g_k = (F_n, F_n, F_\alpha, -F_n F_n, -F_n F_\alpha, -F_n F_\alpha, F_n F_n F_\alpha),$$
$$a_k = (\eta_n, \eta_n, \eta_\alpha, 2\eta_n, (\eta_n+\eta_\alpha), (\eta_n+\eta_\alpha), (2\eta_n+\eta_\alpha)),$$

$$b_k = \left(\frac{4}{9}\eta_n, \frac{4}{9}\eta_n, \frac{1}{9}\eta_\alpha, \frac{8}{9}\eta_n, \left(\frac{4}{9}\eta_n + \frac{1}{9}\eta_\alpha\right), \left(\frac{4}{9}\eta_n + \frac{1}{9}\eta_\alpha\right), \left(\frac{8}{9}\eta_n + \frac{1}{9}\eta_\alpha\right)\right),$$

$$c_k = \left(\frac{1}{4}\eta_n, \frac{1}{4}\eta_n, 0, \frac{1}{2}\eta_n, \frac{1}{4}\eta_n, \frac{1}{4}\eta_n, \frac{1}{2}\eta_n\right),$$

$$d_k = \left(\frac{4}{3}\eta_n, \frac{4}{3}\eta_n, -\frac{2}{3}\eta_\alpha, \frac{8}{3}\eta_n, \left(\frac{4}{3}\eta_n - \frac{2}{3}\eta_\alpha\right), \left(\frac{4}{3}\eta_n - \frac{2}{3}\eta_\alpha\right), \left(\frac{8}{3}\eta_n - \frac{2}{3}\eta_\alpha\right)\right)$$

$$e_k = (\eta_n, -\eta_n, 0, 0, \eta_n, -\eta_n, 0),$$

$$h_k = \left(-\frac{2}{3}\eta_n, \frac{2}{3}\eta_n, 0, 0, -\frac{2}{3}\eta_n, \frac{2}{3}\eta_n, 0\right).$$

The summation over $k$ in the formula of Eq. (14) denotes the summation over the scattering order $k = 1$–3 – single collisions, $k = 4$–6 – double collisions, $k = 7$ – thriple collision.

Substituting the WFs of Eqs. (7), (8) in the formula of Eq. (14) we can write the matrix element:

$$M_{if}(\mathbf{q}) = M0_{if}(\mathbf{q}) + M1_{if}(\mathbf{q}) + M2_{if}(\mathbf{q}), \tag{15}$$

where

$$M0_{if}(\mathbf{q}_\perp) = \sum_{M_s M_s'} \frac{ik}{2\pi} \int d\boldsymbol{\rho} \langle \Psi_{000} | \Omega | \Psi_{000} \rangle, \tag{16}$$

$$M1_{if}(\mathbf{q}_\perp) = \sum_{M_s M_s'} \frac{ik}{2\pi} \int d\boldsymbol{\rho} \langle \Psi_{111} | \Omega | \Psi_{111} \rangle, \tag{17}$$

$$M2_{if}(\mathbf{q}_\perp) = \sum_{M_J M_J'} \frac{ik}{2\pi} \int d\boldsymbol{\rho} \{\langle \Psi_{000} | \Omega | \Psi_{111} \rangle + \langle \Psi_{111} | \Omega | \Psi_{000} \rangle\}. \tag{18}$$

It is important to note that due to this approach (recording of WFs of Eqs. (9), (10) and the operator of Eq. (14) in the form of Gaussian expansion), it is possible to calculate all matrix elements analytically without any simplifications, and therefore without loss of accuracy. The detailed calculations are given in Refs. 22, 23.

## 3.2. *Calculation of the central part of $p^8He$-scattering amplitude*

If we use the density function $\rho(r) = |\Psi(r)|^2$ instead of WF of $^8$He, the calculation is greatly simplified. In this case we do not convert the series of Eq. (2) in the form of Eq. (3), and restrict ourselves with two first terms of the series, because we know that every next term gives contribution to the cross section much lower than the previous one.[10]

The substitution of the multiple scattering series of Eq. (2) (together with of Eq. (5)) in the amplitude of Eq. (1), and its subsequent integrations over the impact parameter $d\boldsymbol{\rho}$ and the momentum transfer in each scattering act $d\mathbf{q}_\mu, \ldots d\mathbf{q}_\nu$, leads to the following result:

$$\tilde{\Omega}(\boldsymbol{\rho},\mathbf{q}) = \frac{2\pi}{ik} f_{pN}(q) \sum_{i=1}^{8} \tilde{\omega}_i - \left(\frac{2\pi}{ik} f_{pN}\left(\frac{q}{2}\right)\right)^2 \sum_{i<j=1}^{8} \tilde{\omega}_{ij} + ..., \tag{19}$$

where

$$\tilde{\omega}_i = \exp(i\mathbf{q}\boldsymbol{\rho}_i), \quad \tilde{\omega}_{ij} = \exp\left(i\frac{\mathbf{q}}{2}(\boldsymbol{\rho}_i + \boldsymbol{\rho}_j)\right)\delta(\boldsymbol{\rho}_i - \boldsymbol{\rho}_j). \tag{20}$$

Because, we have densities instead of WF, so the matrix element of Eq. (1) can be written as follows

$$M_{if}(\mathbf{q}) = \frac{ik}{2\pi} \int |\Psi(r)|^2 \tilde{\Omega}(\boldsymbol{\rho},\mathbf{q}) d\mathbf{r}. \tag{21}$$

The matrix element of single collisions, considering the first formula of Eq. (20), can be written as

$$M_{if}^{(1)}(\mathbf{q}) = f_{pN}(q) \sum_{i=1}^{8} \int |\Psi(r)|^2 \exp(i\mathbf{q}\boldsymbol{\rho}_i) d\mathbf{r}. \tag{22}$$

If $\boldsymbol{\rho}_i = \mathbf{r}$, it is possible to expand $\exp(i\mathbf{q}\mathbf{r})$ into series by Bessel functions

$$\exp(i\mathbf{q}\mathbf{r}) = 4\pi \sum_{\lambda=0}^{\infty} \sum_{\mu=-\lambda}^{\lambda} (i)^{\lambda} \sqrt{\frac{\pi}{2qr_\nu}} J_{\lambda+\frac{1}{2}}(qr) Y_{\lambda\mu}(\Omega_q) Y_{\lambda\mu}(\Omega_r), \tag{23}$$

and calculate the integral of Eq. (22) in the spherical system of coordinates. Then, we can obtain the following for single scattering based on 6 $pn$-collisions and 2 $pp$-collisions:

$$M_{if}^{(1)}(q) = \frac{1}{\sqrt{4\pi}} \sqrt{\frac{\pi}{2q}} \{ 6f_{pn}(q) \int_0^\infty |\Psi_{pn}(r)|^2 J_{1/2}(qr) r^{3/2} dr + 2f_{pp}(q) \int_0^\infty |\Psi_{pp}(r)|^2 J_{1/2}(qr) r^{3/2} dr \}. \tag{24}$$

The similar calculations for double scattering with the second formula of Eq. (20) (based on 15 $pn$-collisions, 13 $pp$-collisions) will result in:

$$M_{if}^{(2)}(q) = \frac{1}{\sqrt{4\pi}} \sqrt{\frac{\pi}{2q}} \{ 15 f_{pn}^2(q/2) \int_0^\infty |\Psi_{pn}(r)|^2 J_{1/2}(qr) r^{3/2} dr + 13 f_{pp}^2(q/2) \int_0^\infty |\Psi_{pp}(r)|^2 J_{1/2}(qr) r^{3/2} dr \}. \tag{25}$$

The matrix element based on the first and second order of collisions:

$$M_{if}(q) = M_{if}^{(1)}(q) - M_{if}^{(2)}(q). \tag{26}$$

The "minus" sign appears because the series of multiple scattering of Eq. (2) is alternating.

3.3. *Accounting for the spin-orbit interaction*

To calculate the polarization characteristics, the elementary amplitude shall take into account not only the central $f^c_{pN}(q)$ but also the spin-orbit $f^s_{pN}(q)$ term, so the amplitude can be written as:

$$f_{pN}(q) = f^c_{pN}(q) + f^s_{pN}(q)\boldsymbol{\sigma}\cdot\mathbf{n}. \tag{27}$$

The spin part of nucleon-nucleon amplitude is parametrized by the following standard way:

$$f^s_{pN} = \frac{k\sigma^s_{pN}}{4\pi} qD_s(i+\varepsilon^s_{pN})\exp\left(-\beta^s_{pN}q^2/2\right). \tag{28}$$

Let's write the matrix element of scattering based on spin dependence

$$M_{if}(q) = M^c_{if}(q) + M^s_{if}(q), \tag{29}$$

where $M^c_{if}(q)$ is central, $M^s_{if}(q)$ is spin-orbit part of the matrix element.

Lets' emphasize the key moments of the spin matrix element derivation, which can be written in the following way:

$$M^s_{if}(\mathbf{q}) = \sum_{M_s M'_s}\left\langle \chi_{\frac{1}{2}M_s}\left|\boldsymbol{\sigma}\cdot\mathbf{n}\right|\chi_{\frac{1}{2}M'_s}\right\rangle \sum_{M_J M'_J}\frac{ik}{2\pi}\int d\boldsymbol{\rho}\exp(i\mathbf{q}\boldsymbol{\rho})\left\langle \Psi^{JM_J}_f\left|\Omega^s\right|\Psi^{JM'_J}_i\right\rangle, \tag{30}$$

where $\chi_{\frac{1}{2}M_s}$ is the spin function, the rest symbols are the same as in the formula of Eq. (1). The spin part of multiple-scattering operator can be written similarly to of Eq. (2):

$$\Omega^s = 1 - \prod_{j=1}^A\left(1-\omega^s_j(\boldsymbol{\rho}-\boldsymbol{\rho}_j)\right) = \sum_{j=1}^A\omega^s_j + \sum_{j\langle\mu}\omega^s_j\omega^s_\mu - \sum_{j\langle\mu\langle\eta}\omega^s_j\omega^s_\mu\omega^s_\eta + ...(-1)^{A-1}\omega^s_1\omega^s_2...\omega^s_A, \tag{31}$$

where $\omega^s_j$ is the profile function, depending on elementary $f^s_{pN}(q)$-amplitude

$$\omega^s_j(\boldsymbol{\rho}-\boldsymbol{\rho}_j) = \frac{1}{2\pi ik}\int d\mathbf{q}\exp\left[-i\mathbf{q}(\boldsymbol{\rho}-\boldsymbol{\rho}_j)\right]f^s_{pN}(q). \tag{32}$$

Just as in the calculation of the central part matrix element, we are limited by the first and second multiplicities of collisions in the $\Omega^s$ operator of Eq. (31).

When calculating the spin matrix elements, we need a system of mutually perpendicular basis vectors $\mathbf{n}, \mathbf{p}, \mathbf{q}$. They are connected with each other and with $\mathbf{k}$ and $\mathbf{k}'$ impulses by the following relations:

$$\mathbf{n} = \mathbf{k} \times \mathbf{k}' = \mathbf{p} \times \mathbf{q}, \quad \mathbf{p} = \mathbf{k} + \mathbf{k}', \quad \mathbf{q} = \mathbf{k} - \mathbf{k}'. \tag{33}$$

$$\boldsymbol{\sigma} \cdot \mathbf{n} = \frac{\mathbf{q} \cdot \mathbf{k} \times \boldsymbol{\sigma}}{k^2 \sin \theta}, \tag{34}$$

where $\sin \theta \approx q/k$. The spin matrix element can be calculated using the cyclic components of unit vector $\mathbf{n}_{-\mu}$:[35]

$$\left\langle \chi_{\frac{1}{2}M_{s'}} \middle| \boldsymbol{\sigma} \cdot \mathbf{n} \middle| \chi_{\frac{1}{2}M_{s'}} \right\rangle = (-1)^{1-\mu} \sqrt{3} \left\langle 1\mu \frac{1}{2} M_{s'} \middle| \frac{1}{2} M_s \right\rangle \mathbf{n}_{-\mu}, \tag{35}$$

$$\mathbf{n}_{-\mu} = \sqrt{\frac{4\pi}{3}} Y_{1-\mu}(\mathbf{n}),$$

$$Y_{1-\mu}(\mathbf{n}) = \begin{cases} Y_{11}(\theta_n, \varphi_n) = -\frac{1}{2}\sqrt{\frac{3}{2\pi}} \exp(i\varphi_n) \\ Y_{10}(\theta_n, \varphi_n) = 0 \\ Y_{1-1}(\theta_n, \varphi_n) = \frac{1}{2}\sqrt{\frac{3}{2\pi}} \exp(-i\varphi_n) \end{cases}. \tag{36}$$

The last result of Eq. (36) is obtained at $\theta_n = \frac{\pi}{2}$ as a result of $\mathbf{n}$ perpendicular to the plane of $\mathbf{k}$ and $\mathbf{k}'$ vectors and, as seen from the first of the relations of Eq. (33). We should note that the spin part of the amplitude at elastic scattering is non-zero only for the transitions that change the spin projection; for the transitions without changing the spin projection the matrix elements equal zero.

The DCS is the square modulus of the matrix element:

$$\frac{d\sigma}{d\Omega} = \frac{1}{2J+1} \sum_{M_J M_{J'}} \left[ \left| M_{if}^c(\mathbf{q}) \right|^2 + \left| M_{if}^s(\mathbf{q}) \right|^2 \right]. \tag{37}$$

It is known that the contribution of the spin-orbit interaction in the DCS is small, the first term provides the main contribution.

The analyzing power, characterizing the dependence of the scattering cross section from the spin direction of the incident proton, is expressed through the matrix element as follows:

$$A_y = \frac{2 \operatorname{Re}\left[ M_{if}^c(\mathbf{q}) M_{if}^{s*}(\mathbf{q}) \right]}{d\sigma/d\Omega}, \tag{38}$$

and it is impossible not to use the spin-orbit matrix element for its calculation.

## 4. Results and discussion

The DCSs and $A_y$ of the $p^6$He and $p^8$He elastic scattering at $E \sim 70$ and 700 MeV/nucleon are calculated according to the formulas derived in the previous section. We compare them with available experimental data and calculations either in the Glauber or in other formalisms. The results are shown in Figs. 1−4.

Fig. 1*a* shows the dependence of DCSs of $p^6$He scattering from model WFs, calculated with different potentials of intercluster interactions (see Table) for the energy $E = 71$ MeV/nucleon. The curves *1, 2* and *3* were calculated with WFs in the models 1 and 2 Refs. 18, 19 and the shell model. The experimental data are from Refs. 6, 7.

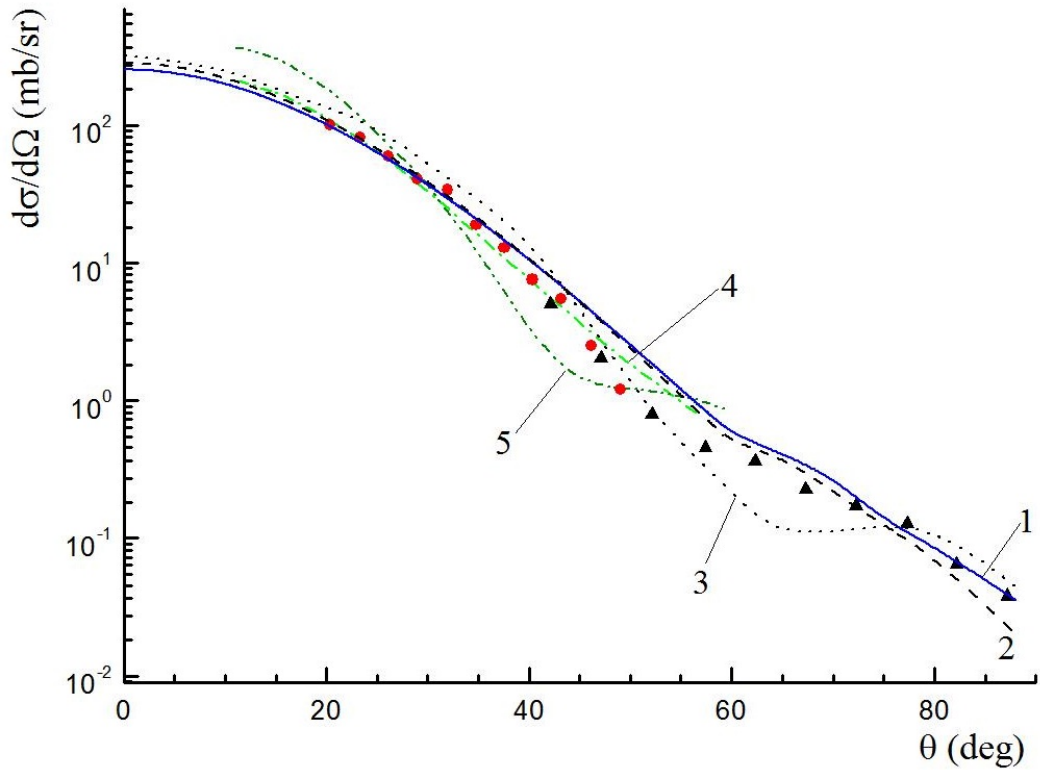

Fig. 1a. Differential cross sections of the $p^6$He elastic scattering at $E = 71$ MeV/nucleon. Experimental data are from Refs. 6, 7. Explanation is given in the text.

The comparison of calculations and experiment shows that for small scattering angles ($\theta < 38°$) DCSs with three-body WFs is in good agreement with experimental data, for the average angles ($\theta > 38°$) the calculated curves are higher than the experimental points, for angles $\theta > 60°$ the agreement with the experiment may be random, since the Glauber approximation is limited by front scattering angles. We have calculated DCSs up to $\theta = 88°$ because new experimental data appeared in Refs. 8, 9. The DCS calculated with the shell WF [20]: $\Psi^{JM_J} = 0.973 \{2\}\ ^{31}S + 0.23 \{11\}\ ^{33}P$ (curve *3*) is worse than with three-body, it describes the experiment both at small and at large angles.

Variation in the description of DCSs with different model WFs is related with their behavior inside the nucleus and the periphery. In the case of small-angle scattering the momentum transfer is small (at $\theta = 2°$, $q = 11$ MeV/c) and only the peripheral region of the nucleus (i.e., asymptotics of WF) can be probed. In the three-body WF the asymptotics is longer than that of the shell WF, which drops quickly and does not convey the real behavior of the nuclear WF. At large scattering angles the momentum transfer increases (reaching the value $q = 217$ MeV/c at $\theta = 40°$), there is more interaction of particles in the interior of the nucleus, where the effects of particle correlations (which actually distinguish one model from another) are more severe and we observe the different behavior of the angular distributions.

For comparison with the results of our calculations we provide the results from Ref. 36, where DCSs are calculated in the folding-model within HEA (High Energy Approximation) with three different densities for $^6$He: LSSM (curve *4*), COSMA (curve *5*) and the density used by Tanihata (it is close to COSMA and not shown in the Figure). In Ref. 36 the microscopic OP was used within HEA with real and imaginary parts and taking into account the spin-orbit interaction. The figure shows that the calculation with COSMA density unsatisfactorily describes the experimental data at all angles. The best agreement with experiment is achieved with LSSM density. It was derived in the full $4\hbar\omega$ shell model using the basic Woods-Saxon one-particle WF with realistic exponential asymptotic behavior. The physical reason for preference of LSSM density is in its more extensive exponential asymptotics, compared to the COSMA density, where WFs are calculated in the potential of harmonic oscillator and have the Gaussian form.

The variation of OP parameters will help to adjust the DCS in all angular range (e.g., set B in Ref. 8), because there are no limits in the optical model for scattering at large angles. These curves are not shown in the figure, because they are very close to the experimental points.

The Fig. 1b shows the DCS of the $p^8$He scattering at the $E = 72$ MeV/nucleon. Our calculation in the Glauber model with LSSM density distribution function from Ref. 21 is shown by the curve *1*. For comparison, we show the curves *2*, *3* from Ref. 37 and *4* from Ref. 38. The curve *2* is calculated in the optical model with WF in LSSM without special adjustment of the potential, the curve *3* − with adjustment of the potential, which includes renormalization of different OP parts. The curve *4* is the best result obtained in approximation of the relativistic mean field (RMF) approach[38] with the new Lagrangian density FSU Gold, which involves the self-coupling vector-isoscalar mesons, as well as coupling between the vector-isoscalar meson and the vector-isovector meson.

These examples show that the DCSs of both $p^6$He and $p^8$He scattering is equally well-described either in the optical model with the phenomenological potential[8,27,37] and in the folding model (cluster and nucleon, Refs. 8, 15, 16, 21) with various functions of density distribution. Naturally, our calculation in the Glauber approximation at such low energy (tens of MeV/nucleon) can not compete from accuracy point of view with the calculation in the optical model. However we should note, that many authors have used the Glauber approximation even at lower energies (e.g. at $E = 41$ MeV, Ref. 39), for qualitative assessment of the cross section). In our case, we also see that the calculated DCS is in good agreement with the experiment at not large angles.

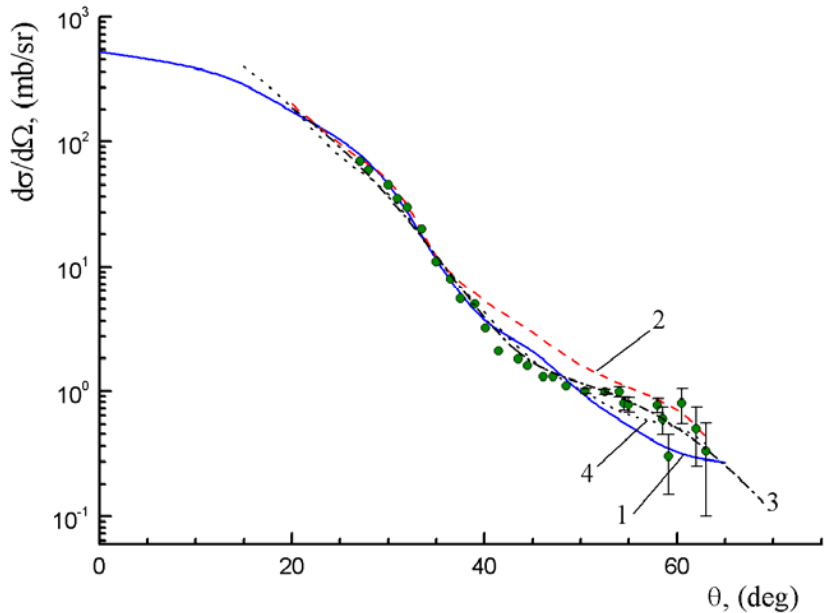

Fig. 1b. Differential cross sections of $p^8$He scattering for the energy $E = 72$ MeV/nucleon. The curve *1* – is our calculation, the curves *2, 3* – is from Ref. 37, the curve *4* – is from Ref. 38. The points are experimental data from Refs. 5, 6.

Fig. 2a shows the same dependence as in Fig. 1*a* for the energy $E = 717$ MeV/nucleon. The curves *1*, *2* and *3* were calculated with WFs in models 1, 2 and shell model. The experimental data from Ref. 1 (DCS is measured up to $\theta < 12°$) and from Ref. 3 (DCS is measured up to $\theta \sim 24°$). It can be seen that the cross sections, calculated by us with the three-body $\alpha-n-n$ WFs, are very close to each other (curves *1* and *2*), indicating low sensitivity of WFs to various intercluster interaction potentials. The calculation with shell WF differs from cluster WFs, which is especially noticeable in the region of the cross section minimum and at large angles. For comparison with our calculation, we present the calculation from Ref. 39 – curve *4* and from Ref. 40 – curve *5*. The calculations of DCS in these papers are also performed in the Glauber approximation with all multiplicities of scattering. In Ref. 39, the calculation is made with three WFs: multicluster $\alpha-n-n$, dineutron cluster and the shell one. Good quantitative description of experimental data is achieved with $\alpha-n-n$ WF obtained by stochastic variation method in Ref. 41. We should note that although the calculation was made before the experimental data at $\theta > 12°$,[3] it agrees well with the experiment at least until $\theta \sim 20°$ for this energy. The curve *5* is calculated with the Faddeev three-body WF, which leads to correct static characteristics: the energy of two neutrons separation and the rms material radius of $^6$He. Since the ideology of Refs. 39, 40 is close to our own, these curves (*1, 2, 4* and *5*) are merged at small angles in the figure. However, considerable divergence is observed in the calculated curves at $\theta > 20°$, indicating the various contributions of high-impulse components in WF.

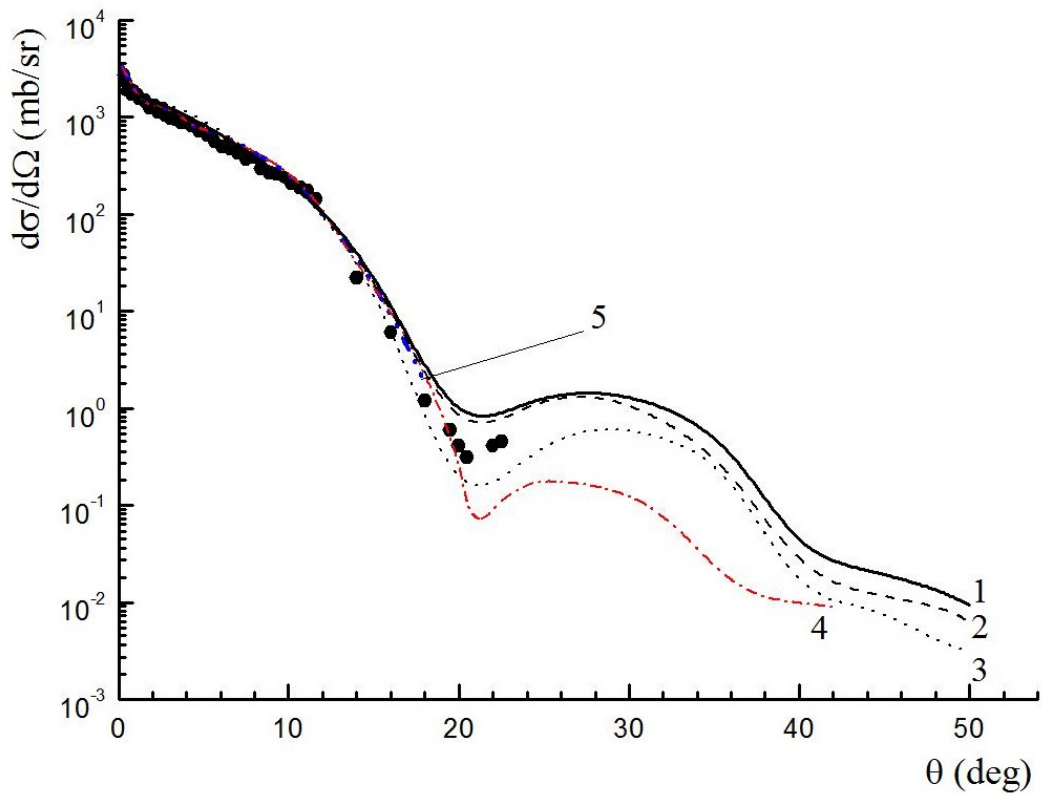

Fig. 2a. Differential cross sections of elastic $p^6$He scattering at $E = 717$ MeV/nucleon. Experimental data are from Refs. 1, 3. Explanation is given in the text.

The completed calculations show that under dependence of DCSs from WFs structure, this dependence is very weak at small scattering angles (where WF asymptotic behavior plays greater role): all curves equally describe the cross section up to the angles $\theta < 20°$. In the area of cross section minimum none of the calculated curves is quantitatively consistent with the experimental data, although the minimum in this region is observed in all of them. However, our calculation with three-body WFs overestimates the cross section at $\theta > 18°$, while the calculation with the oscillator-type WF, as well as the curve *4*, is lower than the experimental data.

About low sensitivity of elastic scattering DCS to various density distributions at small scattering angles is written about in Ref. 42, where the DCS of the $p^8$He elastic scattering is calculated by two different methods: JLM and in the eikonal approximation. For small scattering angles the DCS in both approximations are equally described by the experimental data, while for large angles the calculated curves are different from each other; it can be used to find the difference between the core and the skin. "The contribution from the core dominates at large angles. The described tendency confirms that the large momentum transfer in the scattering occur under interaction with more massive $\alpha$-core".[42] However, the authors point out that the difference in the density distribution in the core and the periphery is not very noticeable and model-dependent, and in order to use it for density distribution measurement the precise measurements of cross sections at large angles are required, which is a difficult task due to generally low cross section at large angles.

The Fig. 2b compares DCSs of the $p^6$He scattering at $E = 717$ MeV/nucleon (curve *1* is the same as curve *1* in Fig. 2a) and the $p^8$He scattering (curve 2) at the energy $E = 671$ MeV/nucleon. Experimental data are from Refs. 2 − 4.

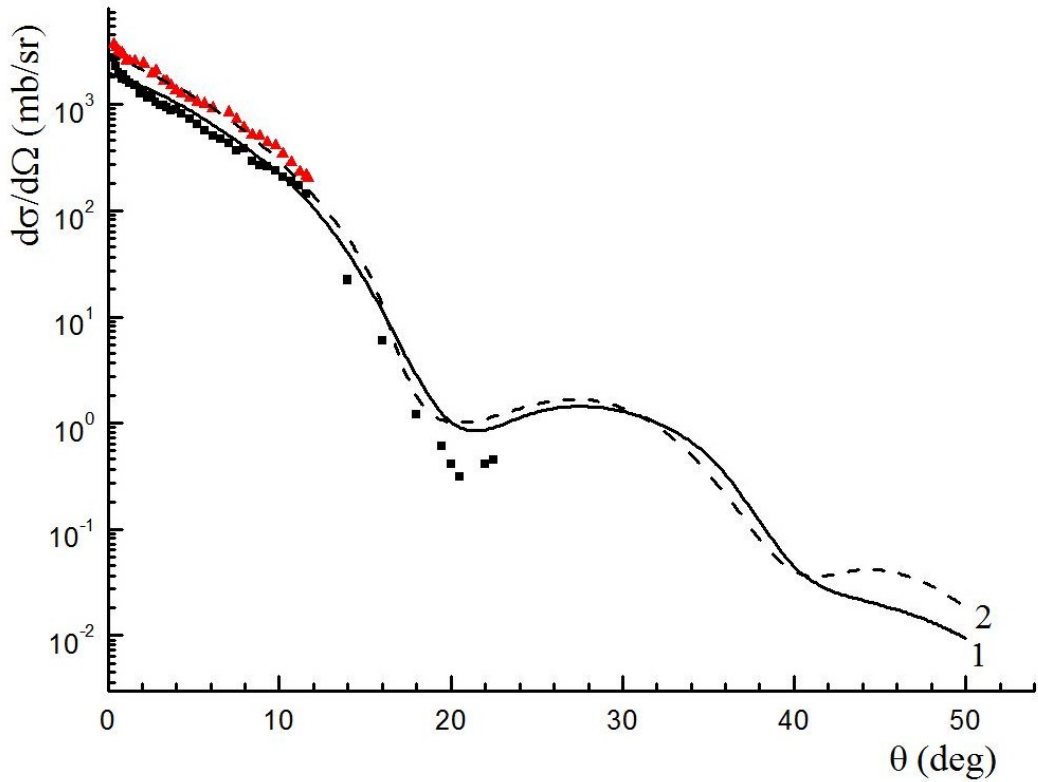

Fig. 2b. Differential cross sections of $p^6$He (curve *1*) and $p^8$He (curve *2*) scattering for the energy $E = 717$ and 671 MeV/nucleon. Experimental data for $p^6$He (squares) are from Refs. 2 − 4, for $p^8$He (triangles) is from Ref. 2 .

The calculated curves are close to each other in the whole angular range, despite the fact that their density distributions are different: three-body WF of $^6$He in model 1 (Table) and LSSM density of $^8$He.[21] This again suggests that the DCS is not very sensitive to the details of the calculation, as noted in the discussion of the results in Fig. 1. However, there are some differences. At small angles DCS of $p^8$He-scattering is more than for $p^6$He. At large angles, where the core contribution dominates, the DCSs for $p^6$He and $p^8$He scattering are also slightly different from each other. While the contribution of core in $^8$He is the same as in $^6$He, the mass effect of valence neutrons appear here, which provides different behavior of DCSs in the range $\theta > 40°$. The comparison of the rms matter and charge radii of $^6$He and $^8$He isotopes, made in,[43] demonstrates the interesting picture: the matter radius of $^8$He is larger than that for $^6$He, while the charge radius is smaller ($R_{ch}^{6He} = 2.068(11)$ fm, $^8$He $R_{ch}^{8He} = 1.929(26)$ fm). It is clear that $R_m^{8He} > R_m^{6He}$ and explained by large numbers of nucleons. The reverse inequality for the charge radii $R_{ch}^{8He} < R_{ch}^{6He}$ is provided by their internal structure. Two extra neutrons in $^6$He are correlated in the way so their finding is more likely on one side of the core (dineutron) than on the opposite side (cigar-like configuration). As a result, the movement of α-core to the correlated pair of neutrons smears the charge distribution on the larger volume. In contrast, in

$^8$He four extra neutrons are distributed more in spherically-symmetric manner in halo and smearing of charge in the core is correspondingly lower, which results in reduction of the charge radius.[43]

The first reliable measurements of the vector analyzing power ($A_y$) in the $p^6$He elastic scattering at 71 MeV/nucleon are obtained recently at the accelerator in RIKEN (Japan).[7-9] Their appearance makes it possible to test different model calculations that take into account the spin-orbit interaction since polarization phenomena in nuclear elastic scattering are the direct indication of the spin-orbit coupling in nuclei.

Fig. 3a shows $A_y$ for $p^6$He scattering at $E$ = 71 MeV/nucleon (*a*) and at $E$ = 717 MeV/nucleon (*b*). The curves *1, 2* and *3* − our calculation with WFs in models 1 and 2 (Table) and shell one. The curves *4−6* are taken from Ref. 9. The experimental data in Fig. 3*a* are from Refs. 8, 9.

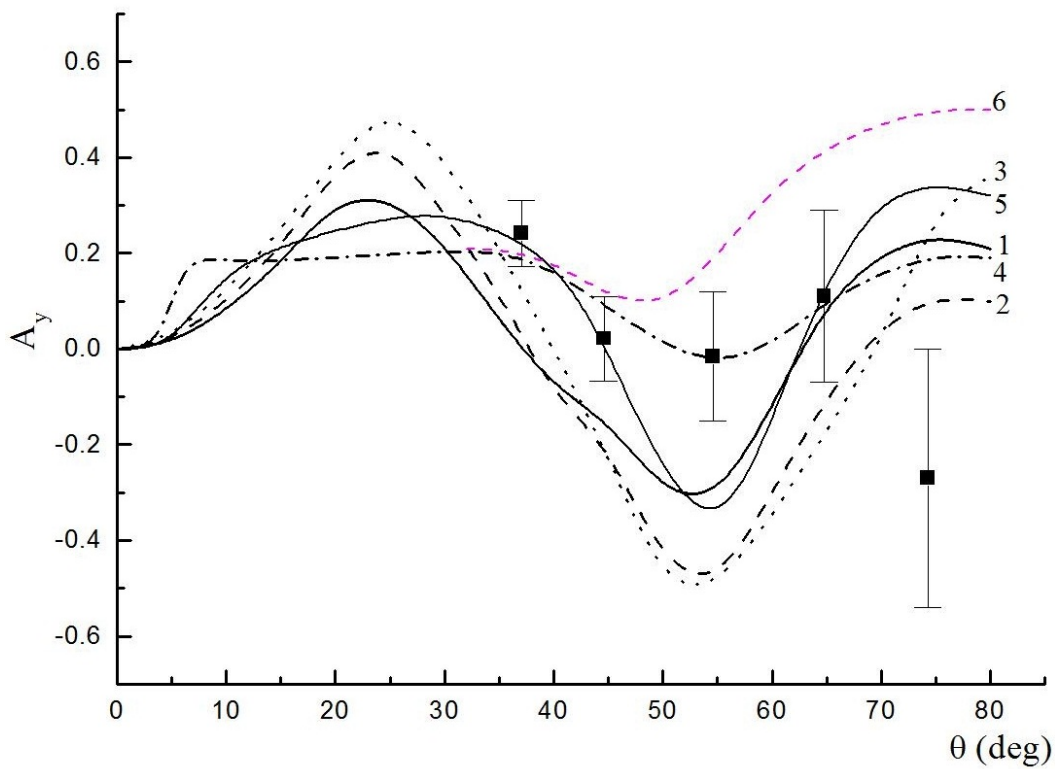

Fig. 3a. Vector analyzing power of $p^6$He scattering at the energy $E$ = 71 MeV/nucleon. The curves *1, 2* and *3* are calculated with WFs in models 1 and 2 (table) and shell one. Curves *4−6* are taken from Ref. 9. Experimental data are from Refs. 8, 9.

Figs. 3a, b calculation at small angles shows approximately the same behavior of $A_y$ for all models WFs (except curve *4*), despite the significant differences of three-body and oscillator WFs of $^6$He. The fact that the calculated curves have the same behavior at small angles (corresponds to small momentum transfer), suggests a weak influence of the peripheral region of the nucleus (scattering on neutron skin or halo) on $A_y$. This is explained by a small spin-orbit contribution to the valence neutrons near zero at small momentum transfers. Though compared to the DCSs (for which all the calculated curves merge at small angles), the polarization characteristics are more different from each other even in the front angles.

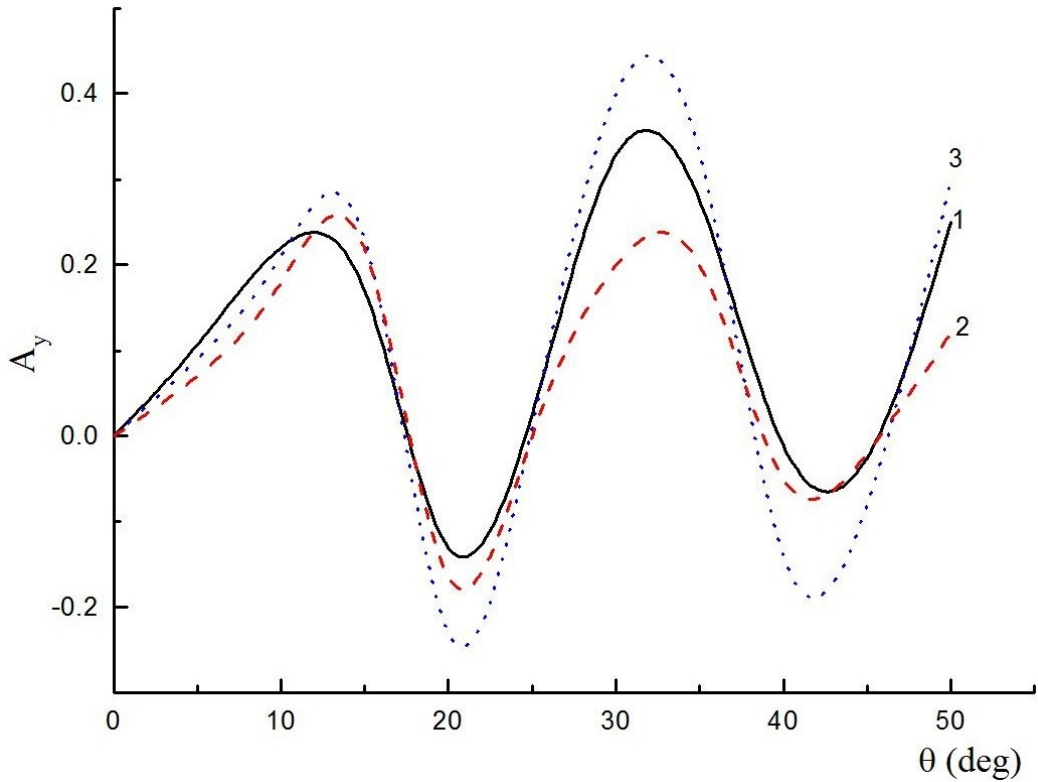

Fig. 3b. The same as in Fig. 3*a* at $E = 717$ MeV/nucleon. The curves *1*, *2* and *3* – calculation with three-body and shell WFs for the $p^6$He scattering.

Increase in the curves amplitude with angle increasing (and momentum transfer) indicates a different contribution of high-impulse WF components. This spread shows considerable sensitivity of the analyzing power to the distribution of nucleons in the central region of the nucleus. Significantly different from each other in oscillations size, all calculated curves reach their maximum and minimum values at the same angles, and with energy increase (from 71 to 717 MeV/nucleon), the number of oscillations increases. At $E = 717$ MeV/nucleon correlation between the minima in DCS (Fig. 2a) and in $A_y$ (Fig. 3b) is observed.

Comparison of our calculations with the experiment in Fig. 3*a* shows only qualitative agreement; in particular, all the curves change the sign from positive to negative at $\theta \sim 40°$. We should note, however, that our calculation[22,23] of $A_y$ was made before the experimental measurements.[8,9]

For comparison, Fig. 3a shows the results of calculation this characteristic from Ref. 9 (curves *4−6*). As mentioned in the introduction, the recent results obtained in Ref. 9 are analyzed there in the optical model with several potentials: phenomenological, folding (cluster and nucleon), non-local in full microscopic model with three sets of single-particle WFs (Woods-Saxon with halo, without halo and with oscillator one). It was possible to provide quantitative description of the experimental data only with the phenomenological OP with special fitting of the parameters (set B, Ref. 9) (curve *4*). The calculation in the cluster folding-model with increased values $r_0$ and $a$ (CF-2, Ref. 9) is in only qualitative agreement with the experiment (curve *5*). Other calculations do not result in satisfactory description of $A_y$. The example is curve *6*, calculated in the microscopic model with Woods-Saxon WF with halo.

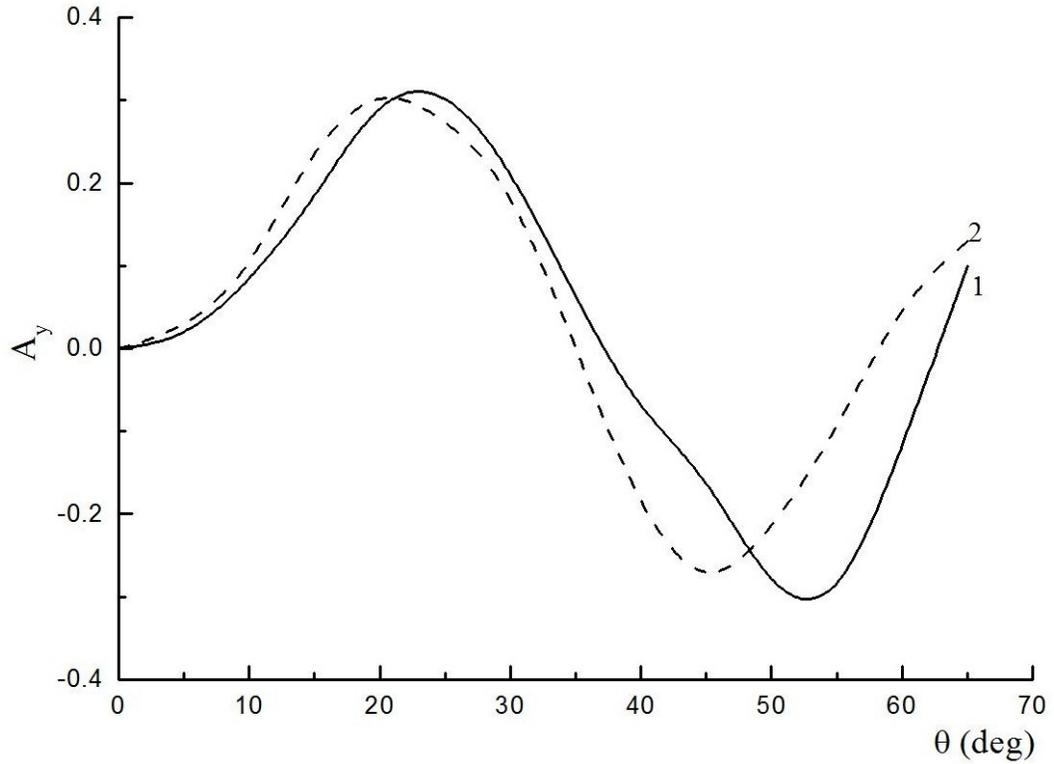

Fig. 4a. Vector analyzing power of the $p^6$He (curve *1*) and $p^8$He (curve *2*) scattering at the energy $E = 71$ and 72 MeV/nucleon.

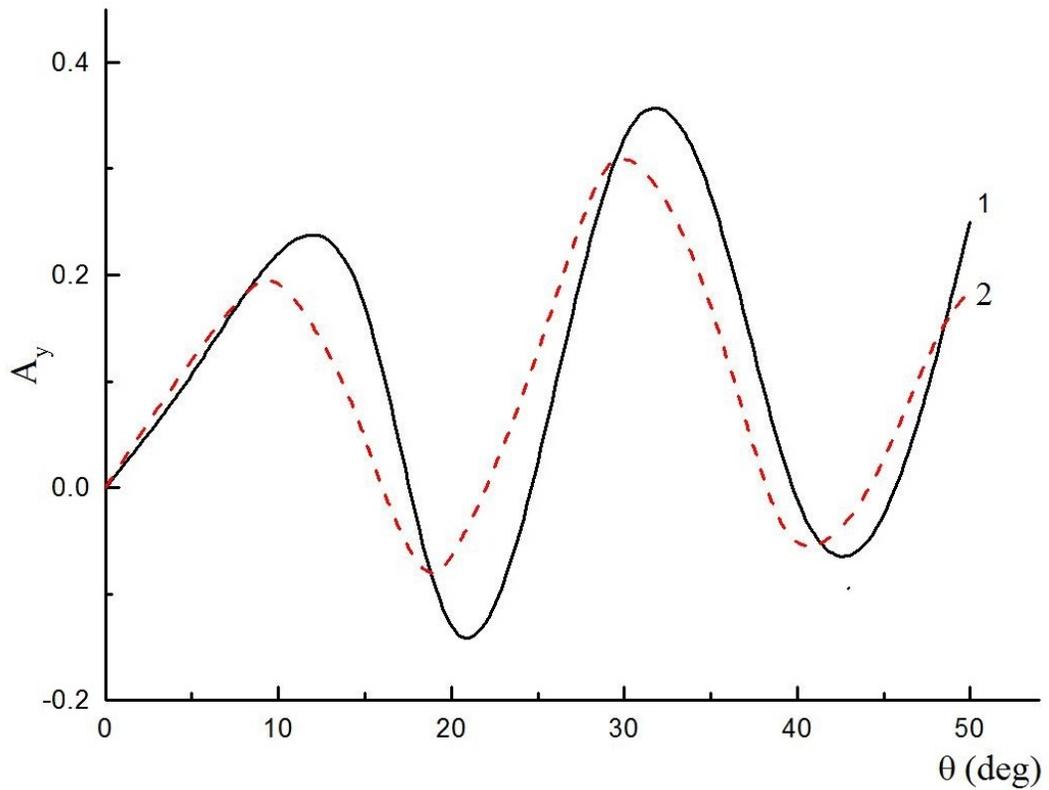

Fig. 4b. The same as in Fig. 4*a* at $E = 717$ and 671 MeV/nucleon.

Fig. 4a, b show a comparison of $A_y$ of $p^6$He (curve *1* is the same as curve *1* in Fig. 3a, b) and $p^8$He (curve *2*) scattering at ~ 70 and ~ 700 MeV/nucleon. It is seen from the figures that the analyzing powers for $p^{6,8}$He are very similar. However, the mass effect of $^8$He valence neutrons is that $A_y$ for $p^8$He is slightly shifted to the region of smaller angles compared to $p^6$He at both energies.

Comparison of DCSs and analyzing powers of $p^{4,6,8}$He and $p^6$Li scattering, made in Ref. 9, revealed interesting features of their behavior. In the angular range $\theta = 20-88°$ DCSs of $p^6$He, $p^8$He and $p^6$Li scattering are similar to each other (and even merge), while DCS of $p^4$He-scattering decreases much more slowly (at $\theta = 88°$ it is larger about 1.5 orders of magnitude than $p^6$He, $p^8$He). The vector analyzing powers for $p^6$He and $p^4$He have the same behavior (changing the sign from positive to negative at $\theta \sim 60°$), whereas $A_y$ for $p^6$Li scattering is positive and increases smoothly without changing the sign. And if the DCSs for proton scattering from these isotopes are described rather accurately in all theoretical approaches (Glauber approximation, high-energy, optical model with phenomenological and folding-potentials), it was much more complicated to reproduce the analyzing power for $p^6$He scattering. Thus, the negative values of this characteristic contradict to most predictions of the optical model. And only the special adjustment of OP parameters provides $A_y$ qualitative description.

## 5. Conclusion

The series of DCSs calculations of protons elastic scattering on nuclei $^6$He and $^8$He are provided here. We used the WFs, obtained within modern three-body nuclear models: $\alpha-n-n$ (for $^6$He) with realistic potentials of intercluster interactions and with the function of density distribution in LSSM (for $^8$He). Expansion of Glauber operator into a series of multiple scattering in the form well adaptable to the picture of weakly bound clusters in halo nuclei, is used for the $p^6$He scattering, the DCS calculation for the $p^8$He scattering is made in the approximation of double collisions. It is shown that the diffraction theory adequately reproduces the experimental data at $E \sim 700$ MeV/nucleon and a bit worse at $E \sim 70$ MeV/nucleon for the angles $\theta > 20°$, due to limitations of the theory, not intended for low energies and large scattering angles.

The importance of using three-body WF for $p^6$He scattering is confirmed by the calculation with the shell WF, which inconsistently describes the experimental DCS, both at low and high scattering angles, due to its rapid decrease on the asymptotic behavior and neglecting nucleons correlations in the interior of the nucleus.

Having calculated $A_y$ for the $p^6$He scattering with different model WFs, we have shown that they much stronger (than DCS) depend on selected WF of the target nucleus in the whole angular range, except for the small-angle scattering, indicating the weak influence of the peripheral region of the nucleus on $A_y$.

Comparison of $A_y$ of the $p^6$He and $p^8$He scattering shows that they are close to each other at both energies, but the mass effect of valence neutrons is that maxima and minima of the $p^8$He scattering curve are shifted to small angles region.

Comparison with the calculations results, made in other approximations (HEA, optical model) and with different model WFs, demonstrated the accuracy of Glauber approximation comparable to the above and description of the characteristics consistent with the experimental data.

**Acknowledgments**

This work was supported by the Grant Program of the Ministry of Education and Science of the Republic of Kazakhstan 1125/GF.